\documentclass[aps,amsfonts,nofootinbib]{revtex4}
\usepackage{epsfig}
\usepackage{graphicx}
\usepackage{amsmath}
\usepackage{amsbsy}
\begin{document}
\newcommand{\ee}{\end{equation}}
\newcommand{\bb}{\begin{equation}}
\newcommand{\eqb}{\begin{eqnarray}}
\newcommand{\eqf}{\end{eqnarray}}
\def\sigmavec{\mbox{\boldmath$\sigma$}}
\def\x{\mathbf{x}}
\def\p{\mathbf{p}}
\def\ho{{\mbox{\tiny{HO}}}}
\def\sc{\scriptscriptstyle}
\newcommand{\1}{{\'{\i}}}
\def\sigmavec{\mbox{\boldmath$\sigma$}}
\def\nablavec{\mbox{\boldmath$\nabla$}}

\title{Bose-Einstein  condensation   theory  for  any   integer  spin:
  approach based in noncommutative quantum mechanics} 
\author{J. Gamboa}
  \email{jgamboa55@gmail.com}
\affiliation{Departamento de  F\'{\i}sica, Universidad de  Santiago de
  Chile, Casilla 307, Santiago, Chile}
\author{F. M\'endez}
  \email{fernando.mendez.f@gmail.com}
\affiliation{Departamento de  F\'{\i}sica, Universidad de  Santiago de
  Chile, Casilla 307, Santiago, Chile}

\begin{abstract}
A  Bose-Einstein  condensation  theory  for  any  integer  spin  using
noncommutative quantum mechanics methods is considered.  The effective
potential   is   derived   as    a   multipolar   expansion   in   the
non-commutativity  parameter  ($\theta$)   and,  at  second  order  in
$\theta$,   our  procedure  yields   to  the   standard  dipole-dipole
interaction  with   $\theta^2$  playing  the  role   of  the  strength
interaction  parameter.   The  generalized  Gross-Pitaevskii  equation
containing  non-local dipolar  contributions is  found. For  $^{52}$Cr isotopes 
$\theta =  C_{dd}/4\pi$ becomes $\sim 10^{-11}$ cm and, thus for this value of $\theta$ one  cannot distinguish 
interactions  coming from  non-commutativity or  those  of dynamical
origin. 
\end{abstract}
%\date{\today}
\maketitle

%%%%%%%%%%%%%%%%%%%%%%%%%%%%%%%%%%%%%%%%%%%%%%%%
%%%%%%%%%%%%%%%%%%%%%%%%%%%%%%%%%%%%%%%%%%%%%%%%
\section{Introduction}
%%%%%%%%%%%%%%%%%%%%%%%%%%%%%%%%%%%%%%%%%%%%%%%%
%%%%%%%%%%%%%%%%%%%%%%%%%%%%%%%%%%%%%%%%%%%%%%%%

Contrarily  to fermions, a  set of  bosons can  occupy a  same quantum
state  as a  consequence of  the Bose-Einstein  statistics. This  is a
purely  quantum effect theorized  by S.   N. Bose  and A.  Einstein in
1924.  During  a very long time  many people thought  that this effect
might be  experimentally unrealizable until it was  discovered in 1995
\cite{nobel}.

By  simplicity,  many research  papers  on Bose-Einstein  condensation
consider  {only}  bosons  with  spin-0  and  {  an  effective  contact
interactions  modeled  by  a  Dirac's  Delta function  and  then,  long
distance interactions  such as} the  dipole-dipole are not  taken into
account. 

Even  though the  dipole-dipole  contributions are  very  small  in
current  experiments,   recently  has  been   shown,  using  $^{52}$Cr
isotopes,  that this interaction  induces explicit  anisotropy effects
which can be experimentally measured \cite{germ1,pfau1,german,german1}.  

Although from microscopic dipole-dipole interactions one can construct
effective potentials, still there are so many open questions concerning
to    the    derivation    of    this    effective    potential    (or
pseudopotential)  in   spite  of  the  intense  work   in  this  field
\cite{liyou}.  

The  importance  of  the  dipole-dipole  interaction can  be  seen  by
considering  two  bosons  with  spins  ${\bf  s}_1$  and  ${\bf  s}_2$
interacting at large distances {through the potential} 
\bb
V =\alpha  \left(\frac{{\bf s}_1. {\bf  s}_2 -3({\bf s}_1.  {\hat {\bf
      r}} )({\bf s}_2. {\hat {\bf r}} )}{r^3}\right), 
\label{dd}
\ee
where {${\bf  r}$ is  the relative position  vector}, ${\hat  {\bf r}}
={\bf r}/r$ and $\alpha$ is the interaction strength \cite{landau1}. The 
interaction  (\ref{dd}) cannot be  eliminated even  in the  dilute gas
limit {{\cite{pfau1} }, and therefore (\ref{dd}) should be included
in the ultracold gas treatment.  

At short  distances, considerations above  are technically implemented
by replacing the two-body interaction simply by a $\delta$-function
{as  the pseudo-potential in  the case  of contact  interactions} and,
therefore, the effective theory  is understood in the Gross-Pitaevskii
approach, {\it i.e} the many-body Hamiltonian becomes 
\bb 
H  = \sum_{i=1}^N  \left(-\frac{1}{2}\nabla^2 +  u({\bf r})  \right) +
\sum_{i>j} W({\bf r}_i -{\bf r}_j). 
\label{mb}
\ee 
with $W({\bf r}_i -{\bf r}_j) = \gamma\,\delta ({\bf r}_i -{\bf r}_j)$
where {  we have set  $m$ and $\hbar$  equals 1. {By other  hand}, the
wave function  in this context  becomes the order parameter  for the
bosonic system.  

Next step in this approach is to solve
\bb
H~\Psi ({\bf r}_1, {\bf r}_2,...,{\bf  r}_N) = E~\Psi ({\bf r}_1, {\bf
  r}_2,...,{\bf r}_N), \label{esc} 
\ee 
where $\Psi ({\bf r}_1, {\bf  r}_2,...,{\bf r}_N)$ is a symmetric wave
function that in the Bose-Einstein phase one can write 
%$\Psi ({\bf r}_1, {\bf r}_2,...,{\bf r}_N)$ 
as  
\bb 
\Psi ({\bf r}_1, {\bf r}_2,...,{\bf r}_N) = \prod_{i=1}^N \psi ({\bf r
}_i ), \label{hartree} 
\ee 
where --of  course-- we have  assumed that all  the bosons are  in the
ground state {and  therefore $ \psi ({\bf r}_i)$  are the one-particle
ground state wave  functions}. The symmetric product (\ref{hartree})
is known as the the Hartree's Ansatz.  
  
Using the Hartree's Ansatz one  can write the energy of this many-body
system as 
\bb 
E_{GP}=  \frac{N}{2} \int  d^3r  \left[ \psi^{*}  ({\bf  r}) \left(  -
  \nabla^2  +  \omega^2 {\bf  r}^2  \right)  \psi  ({\bf r})  +  \gamma
  (N-1)\,(|\psi ({\bf r})|^2)^2 \right], \label{ener0} 
\ee
where 
%in (\ref{ener0}) 
the  one-body potential  $u({\bf r})$  has been  chosen as  a harmonic
oscillator  one  {  as the  trapping  to  confine  the atoms  in  a
ultracold atoms experiment}.

Equation (\ref{ener0}),  {by other hand},  is a direct  consequence of
the  variational  method  and  the  equation  of  motion  obtained  by
minimizing (\ref{ener0}) subject to the normalization condition 
\bb
\int d^3 |\psi ({\bf r}|^2 =1,  \label{suple}
\ee
in this context  { turn out to be}
\bb
\left[ -\frac{1}{2} \nabla^2 +  \frac{1}{2}\omega^2 {\bf r}^2 + \gamma
  (N-1)\, |\psi ({\bf r})|^2 \right] \psi 
({\bf r}) = \mu \psi ({\bf r}), \label{gp0}
\ee 
which  is the  Gross-Pitaevskii  equation and  the chemical  potential
$\mu$ was introduced as a Lagrange multiplier {that takes into account
the condition (\ref{suple})}. 
 
In (\ref{esc}), { as well as in all the discussion that follows it},
except  by the symmetry  of the  wave function,  the spin  effects are
neglected and therefore,  in this approximation $\psi ({\bf  r})$ is a
scalar.   The  purpose  of  the  present  paper will  be  to  look  at
Bose-Einstein  condensation theory  from  the non-commutative  quantum
mechanics point of view as it was formulated in \cite{fglm} and derive
the  Gross-Pitaevskii equation  in this  context. Additionally  we will
present the derivation of  the pseudopotential in this non-commutative
quantum mechanics scenario. 

The paper is organized as follows; in Section II we review briefly our
previous work on NCQM and magnetic dipole interactions, in Section III
we extend  the Hartee's approach to  many body NCQM and  the ideas are
applied to Bose-Einstein condensation and generalized Gross-Pitaevskii
equations are found. Section V contains our conclusions.
 
%%%%%%%%%%%%%%%%%%%%%%%%%%%%%%%%%%%%%%%%%%%%%%%%%%%%%%%%%%%%%%%%%%%
%%%%%%%%%%%%%%%%%%%%%%%%%%%%%%%%%%%%%%%%%%%%%%%%%%%%%%%%%%%%%%%%%%%%
 \section{Noncommutative   quantum  mechanics  and   magnetic  dipolar
   interactions} 
%%%%%%%%%%%%%%%%%%%%%%%%%%%%%%%%%%%%%%%%%%%%%%%%%%%%%%%%%%%%%%%%%%%
%%%%%%%%%%%%%%%%%%%%%%%%%%%%%%%%%%%%%%%%%%%%%%%%%%%%%%%%%%%%%%%%%%%
 
Usually  noncommutative quantum  mechanics of  $N$ particles  in three
dimensions  is  a model  with  a  given  Hamiltonian ${\hat  H}  (\hat
{p}_i,\hat {x}_j)$, with $\{i,j\} =\{1,2\dots, 3N \}$ and deformed
canonical commutators (see e.g \cite{ncqm}), {\it i.e.}  
\eqb 
\left[ {\hat x}_i,{\hat x}_j\right] &=& i \theta_{ij}, \label{nc1}
\\ 
\left[  {\hat p}_i, {\hat p}_j\right] &=& i B_{ij}, \label{nc2} 
\\ 
\left[  {\hat x}_i, {\hat p}_j\right] &=& i \delta_{ij}, \label{nc3}
\eqf 
where $\theta_{ij}$ and $B_{ij}$ are constant $3N\times 3N$ matrices.
   
Although (\ref{nc1})-(\ref{nc3}) can be  also implemented by using the
Moyal  product  in the  phase  space, it  is  more  convenient to  use
commutative variables  instead noncommutative ones.   Technically this
last fact imply to write $\hat p$ and $\hat x$ according to the rules 
\eqb 
{\hat  x}  _i  \to  {\hat   x}  _i  &=&  x_i  +  \frac{\theta_{ij}}{2}
p_j, \label{bopp1} 
\\ 
{\hat   p}  _i   \to  {\hat   p}   _i  &=&   p_i  +   \frac{B_{ij}}{2}
x_j, \label{bopp2} 
\eqf    
where  $x_i$  and $p_i$  are  the  standard  variables satisfying  the
standard canonical commutators, {\it i.e.} 
\eqb 
\left[  {x}_i,{  x}_j\right] &=&  0=  \left[  {  p}_i, {  p}_j\right],
\nonumber 
\\ 
\left[  { x}_i, { p}_j\right] &=& i \delta_{ij}, \nonumber
\eqf 
these rules, sometimes known as  Bopp's shifts, are the starting point
in noncommutative quantum mechanics.   We would like to emphasize that
the spin properties in the conventional Bopp's shifts does not appears
and, therefore, if these effects are incorporated one must to modify 
these rules. 

In  reference \cite{fglm},  for  the  case of  one  particle in  three
dimensions, the spin was introduced by positing the algebra 
\eqb
\left[{\hat x}_i,{\hat x}_j\right]  &=& i\theta^2 \epsilon_{ijk} {\hat
  S}_k, \nonumber 
\\
\left[{\hat     x}_i,{\hat    p}_j\right]    &=&     i    \delta_{ij},
\,\,\,\,\,\,\,\,\,\,\,\,\,\,\,\,\,\,\,\,\,\,\,\left[{\hat    p}_i,{\hat
    p}_j\right ] 
= 0, \label{3}
\\ 
\left[{\hat x}_i,{\hat  s}_j\right] &=& i  \theta \epsilon_{ijk} {\hat
  s}_k, \,\,\,\,\,\,\,\,\,\,\,\, 
\left[{\hat  s}_i,{\hat  s}_j\right] =  i  \epsilon_{ijk} {\hat  s}_k,
\nonumber 
\eqf 
where $\{i,j\}=\{1,2,3\}$  and $\theta$ is a  parameter with dimension
of  length.    Equation  (\ref{3})  is  just  a   deformation  of  the
Heisenberg's algebra similar to the Snyder one \cite{snyder}.  

Following \cite{fglm}  one realize that  the algebra (\ref{3})  can be
explicitly realized in terms of \emph {commutative} variables by means
of the identification 
\eqb
{\hat x}_i  && \rightarrow {\hat x}_i = x_i + \theta S_i, \nonumber
\\
{\hat   p}_i   &&   \rightarrow   {\hat   p}_i  =   p_i:=   -   \imath
\partial_i, \label{6} 
\\
{\hat s}_i && \rightarrow {\hat s}_i = s_i, \nonumber
\eqf
where  $x_i$ and  $p_i$  are now  canonical  operators satisfying  the
Heisenberg's algebra and $S_i$ are spin matrices. Notice the matricial
character of the non-commutative coordinate operators ${\hat x}_i$.

This  simple  observation  implies  that  any  noncommutative  quantum
mechanical system described by the dynamic equation
\bb
\imath \partial_t | \psi(t) \rangle = 
{\hat  H}  ({\hat  p},{\hat  x},{\hat  s})  |\psi(t)\rangle 
= \left[ \frac{1}{2} {\hat p}^2 + {\hat V} ({\hat x})
\right] |\psi(t)\rangle
\label{8}
\ee
can equivalently be  described by the \emph{commutative} Schr\"odinger
equation 
\bb
\imath \partial_t \psi (\mathbf{x},t)= H(p_i, x_i + \theta s_i)\, \psi
(\mathbf{x},t) \,, 
\label{9}
\ee
where $\psi (\mathbf{x},t)$ is a spinor of $2s+1$-components.

In \cite{fglm}  was discussed that  the noncommutative version  of the
harmonic oscillator whose Hamiltonian is 

\eqb
H &=& -\frac{1}{2} \nablavec^2 + \frac{1}{2} {\hat  \x}^2, \nonumber
\\
&=&-\frac{1}{2}  \nablavec^2+  \frac{1}{2}  {\bf  x}^2 +  \theta  {\bf
  x}\,.\,{\bf s} + \theta^2 {\bf s}^2.  \label{99}
\eqf

The factor $ \theta {\bf s}^2$ can be absorbed by a renormalization of
the  ground state  energy  and, therefore,  the  new contribution  due
noncommutativity  is the  magnetic dipolar  interaction $  \theta {\bf
  x}\,.\,{\bf s}$. Of course if one include many particles effects one
should reproduce the dipole-dipole interaction (\ref{dd}), however our
main purpose below will be implement effective interactions.  

%%%%%%%%%%%%%%%%%%%%%%%%%%%%%%%%%%%%%%%%%%%%%%%%
%%%%%%%%%%%%%%%%%%%%%%%%%%%%%%%%%%%%%%%%%%%%%%%%
\section{Approach to Bose-Einstein condensation for any spin} 
%%%%%%%%%%%%%%%%%%%%%%%%%%%%%%%%%%%%%%%%%%%%%%%%
%%%%%%%%%%%%%%%%%%%%%%%%%%%%%%%%%%%%%%%%%%%%%%%%

Let   us   consider   $N$   atoms   described   by   the   Hamiltonian
(\ref{mb}). In  order to  take into account  the spin effects,  we use
the  prescription  in (\ref{6})  in  the  two-body  potential for  the
non-commutative  coordinates  $\hat{{\bf  r}}$  and  $\hat{{\bf  r}}'$
satisfying (\ref{3}), {\it i.e.}
\bb 
W(\hat{{\bf r}} - \hat{{\bf r}}') = W_s(\hat{{\bf r}}-\hat{{\bf r}}') +
W_{\ell}(\hat{{\bf r}} -\hat{{\bf r}}'), 
\ee 
where  $W_s$  and  $W_{\ell}$  are  the contact  and  large  distances
contributions respectively.  

Let   us  now  write   explicitly  both   potentials  up   to  order
$\theta^2$. For the contact interaction we have
\eqb  
W_s (\hat{{\bf  r}} -\hat{{\bf r}}{'})  &=& \gamma \,\delta (  {\bf r}
-{\bf r}{'} + \theta ({\bf s} -{\bf s}{'})), \nonumber 
\\ 
&=& \gamma \,\int d^3 k e^{i {\bf k} \cdot \Delta {\bf r}} e^{i \theta
  {\bf k}\cdot\Delta {\bf s}}, \nonumber 
\\ 
&= & \gamma \,\delta ({\bf r} -{\bf r}{'}) + \theta \gamma \Delta {\bf
  s}   \cdot  \nabla_{\bf   r}  \delta   ({\bf  r}   -{\bf   r}{'})  +
\frac{\gamma\theta^2}{2}  \left( \Delta  {\bf s}\cdot  \nabla_{\bf r}\right)^2
\delta ({\bf r} -{\bf r}{'})+ {\cal O} (\theta^3), 
\eqf 
with $\Delta {\bf s} = {\bf s} - {\bf s}'$. 

The large distances term turn out to be
\eqb 
W_{\ell}({\bf r} -{\bf  r}') &=& \frac{1}{|{\bf r} -{\bf  r}' + \theta
  ({\bf s} -{\bf s}')|}, \nonumber 
\\ 
&=& \frac{1}{|\Delta {\bf r} |} \left[ 1 + \theta \frac{\Delta {\bf s}
    \cdot \Delta {\bf r}}{|\Delta {\bf r} |^2} -\frac 
{\theta^2}{2|\Delta {\bf r}|^2} \left( \Delta {\bf s}^2 - 3 (\Delta {\bf
  s} \cdot {\Delta {\hat { \bf r}}})^2 \right)+ {\cal O} (\theta^3)\right], 
\eqf 
where  we have  defined $\Delta  {\bf r}  = {\bf  r} -  {\bf  r}'$ and
$\Delta {\hat {\bf r}} = \Delta {\bf r}/| \Delta {\bf r}|$.  
 
Since the spin contributions are  dominant at large distances, one can
neglect the purely coulombian contributions, thus the the potential at
large distances becomes 
\bb 
W_{\ell}({\bf  r}  -{\bf  r}')  \approx \theta  \frac{\Delta  {\bf  s}
  \cdot\Delta {\bf r} }{|\Delta {\bf r} |^3}- \frac {\theta^2}{2}
\frac{\left(\Delta  {\bf s}^2 -  3(\Delta {\bf  s} \cdot  \Delta{ \hat
    {\bf r}})^2 \right)}{|\Delta {\bf r}|^3} 
+ \cdots, \label{dd2}
\ee 
which contains the dipole and dipole-dipole contributions and so on. 

At very large distances, of  course there are not contact interactions
and the  dominant term of  the pseudo-potential is  (\ref{dd2}) which,
compared with (\ref{dd}) tells us
\bb
\alpha =\frac{1}{2}\theta^2.   \label{ident}
\ee 
It  is interesting to  note that  the identification  (\ref{ident}) is
similar  to the relationship  between NCQM  and the  Landau's problem,
where  $\theta   =1/B$  corresponds  to  the   magnetic  length.  This
identification  becomes  exact at  the  lowest  Landau  level. In  our
problem at hand, identification  (\ref{ident}) plays a similar role to
the non-commutative  Landau's one and here $\theta$  could be directly
extracted from  known data.  Indeed, following \cite{pfau1}  one find
that
$$
\theta^2=\frac{C_{dd}}{4\pi}=\frac{48 a_0\hbar^2}{m}
$$
where  $a_0$ is the  Bohr's radius  and $m$  is the  $^{52}$Cr isotope
mass. We therefore find 
\begin{equation}
\theta\sim 10^{-11} \mbox{cm},
\end{equation}
which is  the analogous of  the magnetic field in  the 
Landau problem.

%%%%%%%%%%%%%%%%%%%%%%%%%%%%%%%%%%%%%%%%%%%%%%%%%%%%%%%%%%%%%%%%%%%%%
%%%%%%%%%%%%%%%%%%%%%%%%%%%%%%%%%%%%%%%%%%%%%%%%%%%%%%%%%%%%%%%%%%%%%
\subsection{Generalized Gross-Pitaevskii equations}
%%%%%%%%%%%%%%%%%%%%%%%%%%%%%%%%%%%%%%%%%%%%%%%%%%%%%%%%%%%%%%%%%%%%%
%%%%%%%%%%%%%%%%%%%%%%%%%%%%%%%%%%%%%%%%%%%%%%%%%%%%%%%%%%%%%%%%%%%%%

Following the Hartree's approach let us consider the derivation of the
Gross-Pitaevskii equations; in this approach the functional of energy is 
$$
E=\frac{<\Psi|\hat H | \Psi>}{<\Psi|\Psi>}, 
$$
where $\hat{H}$ is the many body Hamiltonian (\ref{mb}) and is assumed
that  the  $N$ bosons  are  in  the ground  state.  In  this case  $E$
becomes\footnote{In what follows we  have changed the normalization of
  the wave function as  $\psi({\bf r})\rightarrow \sqrt{N} \psi({\bf r})$}
\eqb 
E  &=&  \int  d^3r  \left[ \psi^{*}  ({\bf  r}) \left(  -\frac{1}{2}
  \nabla^2  + \frac{1}{2}  \omega^2  \left({\bf r}  +  \theta {\bf  s}
  \right)^2 \right) \psi ({\bf r}) + \frac{\gamma}{2}\, (|\psi ({\bf r})|^2)^2
  \right] \nonumber 
\\
 &+& \frac{\gamma \theta}{2}\,\int d^3 r \,d^3 r' \psi^{*} ({\bf
   r}) \, \psi^{*} ({\bf r}')\, \left (\Delta {\bf s} \cdot 
 \nabla_{\bf r} \delta ({\bf r} - {\bf r}') +
  \theta \left( \Delta {\bf s}\cdot \nabla_{\bf r}\right)^2 \delta ({\bf r}
 -{\bf r}') \right) \psi ({\bf r}') \psi ({\bf r}) \nonumber 
\\ 
  &+& \frac{ \theta}{2}\, \int d^3 r  \,d^3 r^{'} \psi^{*}
({\bf  r})\, \psi^{*}  ({\bf r}')\,  \left (  \frac{\Delta  {\bf s}
  \cdot  \Delta {\bf  r}  }{|\Delta {\bf  r}  |^3} -  \frac{\theta}{2}
\frac{\left(\Delta {\bf s}^2 - 3 (\Delta {\bf s} \cdot \Delta{ \hat{ \bf
    r}})^2  \right)}{|\Delta  {\bf  r}|^3} \right)  \psi  ({\bf
  r}') \psi ({\bf r}).  \label
{ener03}
 \eqf 
 
 Using some identities, equation (\ref{ener03}) can be written as 
\bb
E =  \left( E_{GP} +E_{C}+ E_D \right) + {\cal O} (\theta^3),  \label{ener2}
\ee  

The second contribution is 
\bb 
E_C= \frac{ \gamma   \theta} {2}  \int  d^3   r  \left[  -\nabla
  n\,\cdot \,\psi^*({\bf r}) {\bf s} \psi ({\bf r}) + \left 
(  \nabla n\cdot  {\bf s}\right)  \left( \nabla  \psi^* \cdot  {\bf s}
  \right) \psi ({\bf r}) \right], 
\ee 
where $n = |\psi|^2$ is the density of particles. 

The energy $E_D$, by other hand, can be written 
\bb 
E_D = \frac{ \theta}{2} \int d^3 r \, \psi({\bf r}) U_{DD}
({\bf r})\, \psi ({\bf r}), 
\ee 
with 
\bb 
U_{DD} ({\bf  r}) = \int  d^3 r' \psi^*({\bf r}')  \left[ \frac{\Delta
    {\bf s} \cdot \Delta {\bf r} }{|\Delta {\bf r} |^3}+\frac {\theta}{2}
  \frac{\left(\Delta  {\bf  s}^2  -  3(\Delta  {\bf  s}  \cdot  \Delta
    \hat{{\bf r})}^2 \right)}{|\Delta {\bf r}|^3} \right] \psi ({\bf r'}), 
\ee 
which corresponds to the dipole and dipole-dipole contributions. 
 
However, before  to minimize (\ref{ener2}), one should  note that {\bf
  the depletion}  is very small and,  therefore, we can  leave out all
terms containing derivatives of $n$, thus 
$$ 
E_C \approx 0. 
$$
Finally,   using  this  last   fact,  we   arrive  to   the  following
Gross-Pitaevskii equation 
\bb
\left[   -\frac{1}{2}   \nabla^2   +  \frac{1}{2}\omega^2   \left({\bf
    r}+\theta {\bf s} \right)^2+ \frac{\gamma}{2}\, |\psi ({\bf r})|^2
  + \frac{ \theta}{2} U_{DD}
  ({\bf r} ) \right] \psi ({\bf r}) = \mu \psi ({\bf r}), 
\label{gp221}
\ee 
which, except for the dipole term $ \frac{\Delta {\bf s}\cdot \Delta {\bf
    r} }{|\Delta {\bf r}  |^3}$, is just the Gross-Pitaevskii equation
with the nonlocal interaction discussed in \cite{pfau1}.  

The non-local equation (\ref{gp221}) is complicated to solve. However one could try solve perturbatively this equation by assuming axial symmetry and to adding the non-local term as a perturbation one. In such case at $0$-th order one find non-intercating vortices solutions, however the explicit inclusion of $U_{DD}$ could induce drastic changes.
%%%%%%%%%%%%%%%%%%%%%%%%%%%%%%%%%%%%%%%%%%%%%%%%
%%%%%%%%%%%%%%%%%%%%%%%%%%%%%%%%%%%%%%%%%%%%%%%%
\section{Conclusions} 
%%%%%%%%%%%%%%%%%%%%%%%%%%%%%%%%%%%%%%%%%%%%%%%%
%%%%%%%%%%%%%%%%%%%%%%%%%%%%%%%%%%%%%%%%%%%%%%%%

In this  paper we have  studied the Bose-Einstein condensates  for any
integer  spin  by  using  non-  commutative  quantum  mechanics  as  a
calculation technique. This procedure induces effective dipole and dipole-dipole interactions which coincide with previous calculations \cite{pfau1,liyou}. The comparison with (\ref{dd}) yields to  (\ref{ident})  and, therefore, this identification is similar to the relationship between $\theta$ and the magnetic length in the Landau problem which is another example where the non-commutative geometry is a useful calculation technique. This last fact follows from the identification between $\theta$ and $C_{dd}/4\pi$ which is just the border between non-commutativity effects and dipole-dipole interactions. Only if (\ref{ident}) is holds, then one cannot distinguish interactions coming from non-commutativity or dynamical origin.

\vspace{0.3 cm}

\noindent\underline{Acknowledgements}: 
We would like to thank H. Falomir and J. L\'opez-Sarri\'on by useful discussions.  This  work was  partially supported  by  FONDECYT-Chile grant-1095106,
1060079.

\end{document}